%% file: paper_pds456_v3_unbold.tex
\documentclass[useAMS,usenatbib]{mn2e}
\usepackage{amssymb,amsmath,epsfig,times,longtable,color,graphicx}
\usepackage{hyperref}

\input{journal_defs}

\newcommand{\nustar}{\textit{NuSTAR}}
\newcommand{\suzaku}{{\it Suzaku}}

\newcommand{\xmm}{{\it XMM-Newton}}

\makeatletter
 \def\hlinewd#1{%
   \noalign{\ifnum0=`}\fi\hrule \@height #1 \futurelet
    \reserved@a\@xhline}
\makeatother

\title[PCA of PDS 456]{Using principal component analysis to understand the variability of PDS~456}

\author[M. L. Parker et al.]{M. L. Parker$^{1, 2}$\thanks{Email: mparker@sciops.esa.int},
J. N. Reeves$^{3,4}$,
G. A. Matzeu$^{5}$,
D. J. K. Buisson$^{1}$,
and A. C. Fabian$^{1}$\\
 $^{1}$European Space Astronomy Centre (ESA/ESAC), E-28691 Villanueva de la Ca\~{n}ada, Madrid, Spain\\
  $^2$Institute of Astronomy, Madingley Road, Cambridge, CB3 0HA, UK\\
  $^3$Center for Space Science and Technology, University of Maryland Baltimore County, 1000 Hilltop Circle, Baltimore, MD 21250, USA\\
  $^4$Astrophysics Group, School of Physical and Geographical Sciences, Keele University, Keele, Staffordshire ST5 5BG, UK\\
  $^5$INAF--Osservatorio Astronomico di Brera, Via Bianchi 46, I-23807 Merate (LC), Italy 
}
\date{}

\begin{document}

\maketitle

\begin{abstract}
We present a spectral-variability analysis of the low-redshift quasar PDS~456 using principal component analysis. In the \xmm\ data, we find a strong peak in the first principal component at the energy of the Fe absorption line from the highly blueshifted outflow. This indicates that the absorption feature is more variable than the continuum, and that it is responding to the continuum. We find qualitatively different behaviour in the \suzaku\ data, which is dominated by changes in the column density of neutral absorption. In this case, we find no evidence of the absorption produced by the highly ionized gas being correlated with this variability. Additionally, we perform simulations of the source variability, and demonstrate that PCA can trivially distinguish between outflow variability correlated, anti-correlated, and un-correlated with the continuum flux. Here, the observed anti-correlation between the absorption line equivalent width and the continuum flux may be due to the ionization of the wind responding to the continuum. Finally, we compare our results with those found in the narrow-line Seyfert 1 IRAS~13224--3809. We find that the Fe K UFO feature is sharper and more prominent in PDS~456, but that it lacks the lower energy features from lighter elements found in IRAS~13224--3809, presumably due to differences in ionization.
\end{abstract}

\begin{keywords}
accretion, accretion discs, black hole physics, X-rays: galaxies
\end{keywords}

\section{Introduction}

Outflows from active galactic nuclei (AGN) are a promising candidate for driving AGN feedback \citep[see review by][]{Fabian12_feedback}. Because of the wider opening angle relative to jets, winds from the AGN accretion disk can potentially couple more effectively to galactic gas, regulating the expansion of the host galaxy. The most extreme such outflows are known as ultra-fast outflows (UFOs), and have outflow velocities of 0.03--0.4$c$. These outflows are most frequently detected using high energy Fe\textsc{xxv/xxvi} absorption lines, found in the 7--10~keV energy range \citep[e.g.][]{Tombesi10, Gofford13}. However, this energy band is towards the edge of the detector range of the current generation of CCD detectors, where the sensitivity declines steeply. This means that just detecting these features is very difficult, so there are very few studies on their variability or timing properties.

Principal component analysis (PCA) is a powerful tool for understanding complex datasets, such as the results of large observing campaigns on active galactic nuclei (AGN), which show complex and rapidly variable spectra produced by the interactions of multiple spectral components. We have used this technique extensively to explore the variability of AGN over recent years \citep[e.g.][]{Parker15_pcasample}.
Most recently, in \citet{Parker17_irasvariability}, we demonstrated that the absorption features from the rapidly variable ultra-fast outflow in the extreme narrow-line Seyfert 1 (NLS1) IRAS~13224-3809 produce peaks in simple variability spectra calculated using data from $\sim2$~Ms of observations. Variability peaks due to the Ly$\alpha$ lines of Fe\textsc{xxv/xxvi}, Ca\textsc{xx}, Ar\textsc{xviii}, S\textsc{xvi}, Si\textsc{xiv}, Mg\textsc{xii} and Ne\textsc{x} are clearly present in the combined dataset, with a velocity equal to that found using EPIC-pn and RGS spectroscopy \citep{Parker17_nature, Pinto17}. These variability peaks indicate that the UFO is responding to the source flux, with the equivalent width of the lines dropping as the flux rises. This is most obviously explained by changes in the ionization of outflowing gas in a disk wind, but could potentially also be explained by geometry changes in a scenario where the absorption lines are produced in an absorbing layer on the disk \citep{Gallo11, Gallo13}, In this case, the absorption comes from the outer Thomson depth of the disk, giving a high column density at large viewing angles, and the absorption is applied primarily to the reflection spectrum, so tracks the reflection fraction. A third possibility is that at least part of the continuum variability is driven by changes in column density of the UFO gas, such that electron scattering attenuates the continuum flux as the column rises. This would then produce stronger absorption lines when the observed flux is lower. While having the variability driven by purely column density changes is statistically disfavoured \citep{Pinto17}, it is likely that lower density gas would also be more highly ionized, so linked column density and ionization variability may be able to explain the observed spectral variability.

The prototypical AGN for studying UFOs is the low-redshift ($z=0.184$) quasar PDS~456. This source shows strong and highly variable high-energy Fe absorption features \citep[e.g.][]{Reeves09, Reeves14} from a UFO ($v\sim0.24c$). Unlike IRAS~13224-3809, PDS~456 does not show a trivial correlation between the equivalent width of the Fe absorption line(s) and the source flux, most likely due to complicating neutral absorption. \citet{Matzeu16} demonstrated that the spectral variability within the 2013 \suzaku\ observing campaign contains a mixture of intrinsic source variability and absorption variability, which may obscure any relation between the UFO and the continuum. However, there is a strong correlation between the source flux and the measured velocity of the outflow \citep{Matzeu17}, which may indicate that the wind is driven by radiation pressure. A potential correlation between the source flux and the UFO strength in PDS~456 is also discussed in \citet{Nardini15}, in the context of ionization changes.

Because of the high mass of PDS~456 \citep[$\sim10^{9} M_\odot$,][]{Nardini15} its variability is slow, occurring largely between, rather than during, observations. PCA is therefore an ideal tool for studying the long timescale variability, due to its indifference to the gaps between observations. In this paper, we use PCA to reveal the long-term spectral variability of PDS~456. Specifically, we investigate the linked variability of the X-ray source and outflow, as found in IRAS~13224-3809 by \citet{Parker17_nature} and \citet{Pinto17}.

\section{Observations and Data Reduction}
\label{section_datareduction}

\subsection{X-ray observations}
\label{section_xray_observations}
We use all the available data on PDS~456 from the \xmm\ and \suzaku\ archives. The full lists of observations are given in Tables~\ref{table_xmmobs} and \ref{table_suzakuobs}, respectively. The majority of the \xmm\ observations are from the 2013/14 observing campaign, presented by \citet{Nardini15}, and the majority of the \suzaku\ observations are from the 2013 campaign, presented by \citet{Matzeu16} and compared with the earlier observations in \citet{Matzeu17_flares}.

The \xmm\ data are reduced using the \xmm\ science analysis software (SAS) version 15.0.0. We use only the high signal to noise EPIC-pn data, which we extract using the \textsc{epproc} ftool. We filter the data for background flares, and extract source and background spectra from 40$^{\prime\prime}$ circular regions, avoiding the outer regions of the pn detector, which are affected by high copper background. 

We use the \suzaku\ data from the two front illuminated detectors, XIS0 and XIS3. We re-run the \suzaku\ pipeline with the latest calibration files to generate clean event lists, following the data reduction guide\footnote{\url{http://heasarc.gsfc.nasa.gov/docs/suzaku/analysis/}}. We use \textsc{xselect} to extract spectral products, with 100$^{\prime\prime}$ radius circular source and background regions, avoiding contaminating sources (including calibration sources). We sum the data for XIS0 and XIS3 using \textsc{addascaspec}.

\begin{table}
\caption{Summary of \xmm\ observations used in this work. We show the observation IDs, start dates, exposure times, and the number of 10~ks spectra used in the analysis after the data are reduced and filtered for flaring background.}
\label{table_xmmobs}
\begin{tabular}{l c c r}
\hline
\hline
Obs. ID 	& Start Date & Exposure time (ks) & Intervals\\
\hline
0041160101 	& 2001 Feb 26th 	& 46.5 	& 4\\
0501580101	& 2007 Sept 12th	& 92.4	& 8\\
0501580201	& 2007 Sept 14th	& 89.7	& 8\\
0721010201	& 2013 Aug 27th		& 111.2	& 10\\
0721010301	& 2013 Sept 6th		& 113.5 & 10\\
0721010401	& 2013 Sept 15th	& 120.5 & 11\\
0721010501	& 2013 Sept 20th	& 112.1 & 10\\
0721010601	& 2014 Feb 26th		& 140.8 & 11\\
\hline
\hline
\end{tabular}
\end{table}

\begin{table}
\caption{Summary of \suzaku\ observations used in this work. We show the observation IDs, start dates, exposure times, and the number of 30~ks spectra used in the analysis after the data are reduced and filtered for flaring background.}
\label{table_suzakuobs}
\begin{tabular}{l c c r}
\hline
\hline
Obs. ID 	& Start Date & Exposure time (ks) & Intervals$^{1}$\\
\hline
701056010 	& 2007 Oct 16th 	& 190.6 & 12\\
705041010	& 2011 April 26th	& 125.5	& 8\\
707035010	& 2013 Feb 21st		& 89.7	& 15\\
707035020	& 2013 March 3rd	& 111.2	& 14\\
707035030	& 2013 March 8th	& 113.5 & 9\\
\hline
\hline
\end{tabular}
$^1$Note that the exposure times take into account Earth occultations, and thus are significantly shorter than the 30~ks timestep multiplied by the number of intervals.
\end{table}

\section{Results}
We use the method described fully in \citet{Parker14_mcg6} to decompose the data\footnote{code available from \url{http://www-xray.ast.cam.ac.uk/~mlparker/PCA.tar.gz}}. In brief, the data are sliced into a set of spectra at fixed intervals (10~ks for \xmm , 30~ks for \suzaku ). We convert these spectra to an array of normalised residuals to the mean spectrum, which we decompose using singular value decomposition (SVD). SVD returns a set of orthogonal principal components (PCs), which describe the original data as efficiently as possible, along with their corresponding lightcurves and contributions to the total variance of the data. 
We calculate the errors using the Monte-Carlo method of \citet{Miller08}. We perturb the input spectra with Poisson noise and rerun the analysis 10 times, then calculate the standard deviation in the results.

We analyse the data from the two instruments separately, as they appear to sample different variability regimes \citep[e.g.][]{Matzeu16}. Unless otherwise specified, all plots are in the rest frame of the source.

\subsection{XMM Newton}
\label{subsec_xmmonly}

\begin{figure}
\includegraphics[width=\linewidth]{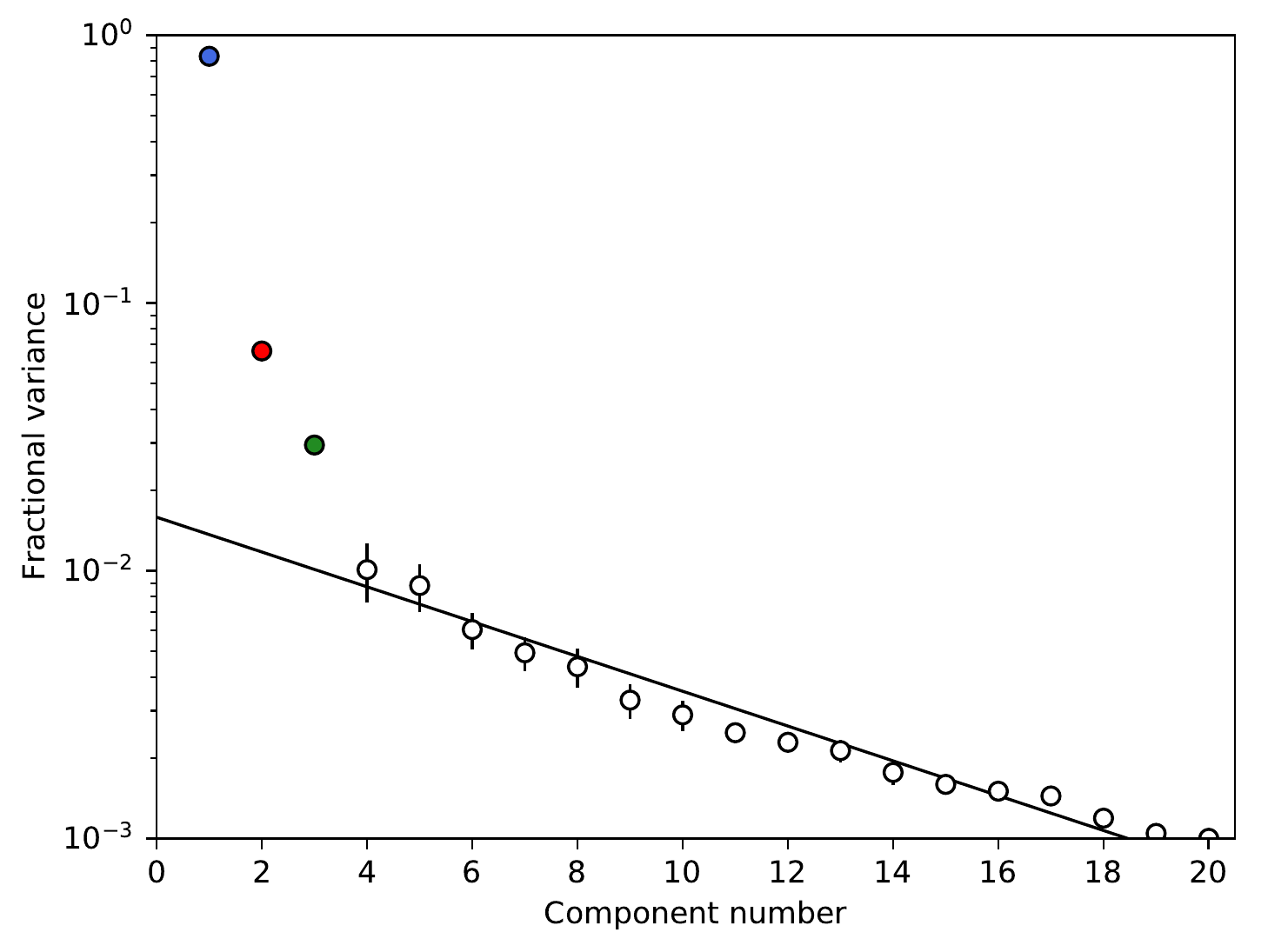}
\caption{LEV diagram showing the fractional variability of each component. The three filled circles are those we identify as highly significant, and which show detailed structure in their spectra. The remainder are consistent with noise. The black line shows a simple geometric decay fit to the higher order components.}
\label{fig_lev}
\end{figure}

In Fig.~\ref{fig_lev} we show the log-eigenvalue (LEV) diagram for the \xmm\ analysis. This shows the fraction of the total variance attributable to each spectral component. Those which are caused by noise should follow a simple geometric decay on this plot. We find that three components are not consistent with such a decay, and focus our analysis on them.

\begin{figure*}
\centering
\includegraphics[width=0.4\linewidth]{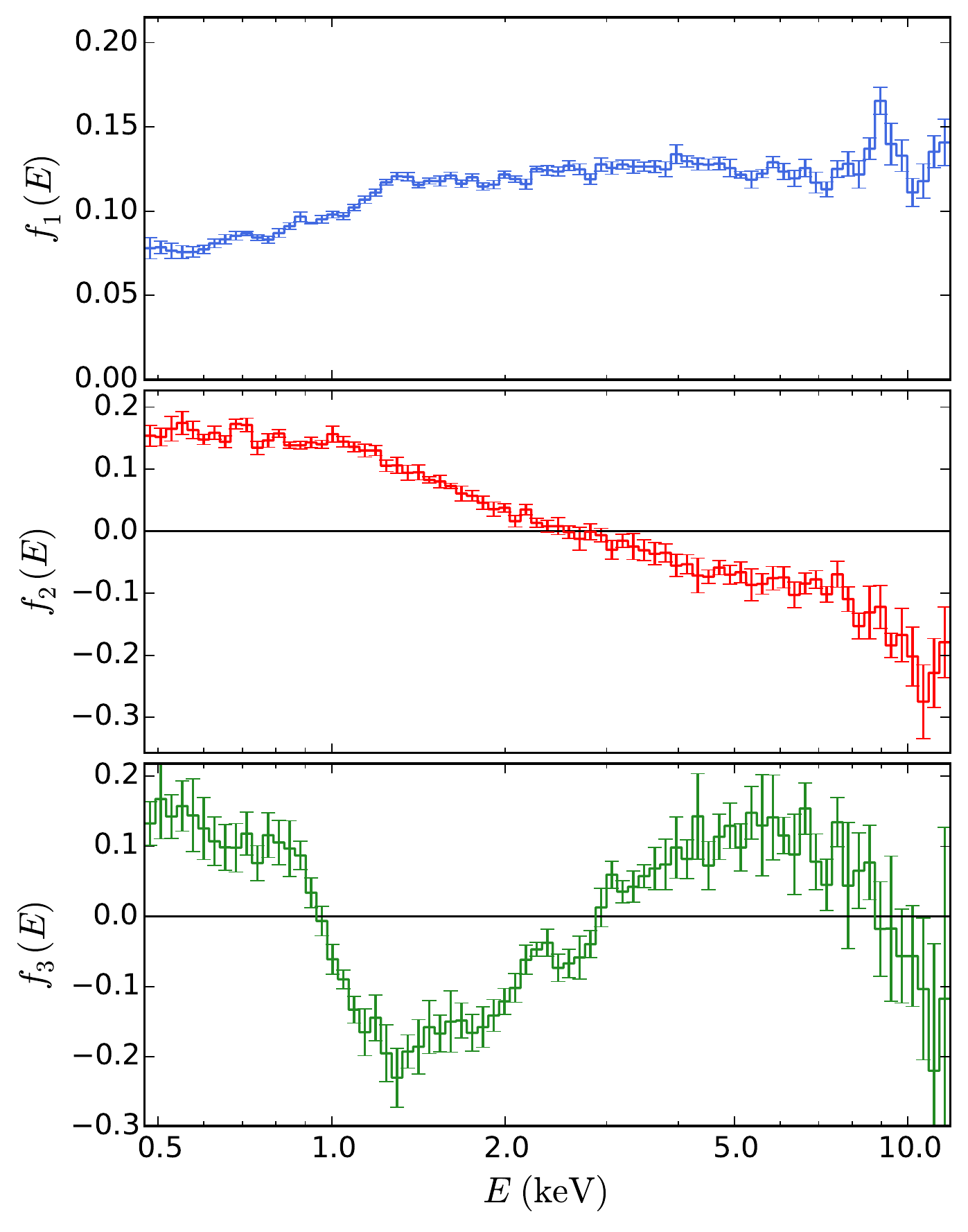}
\includegraphics[width=0.4\linewidth]{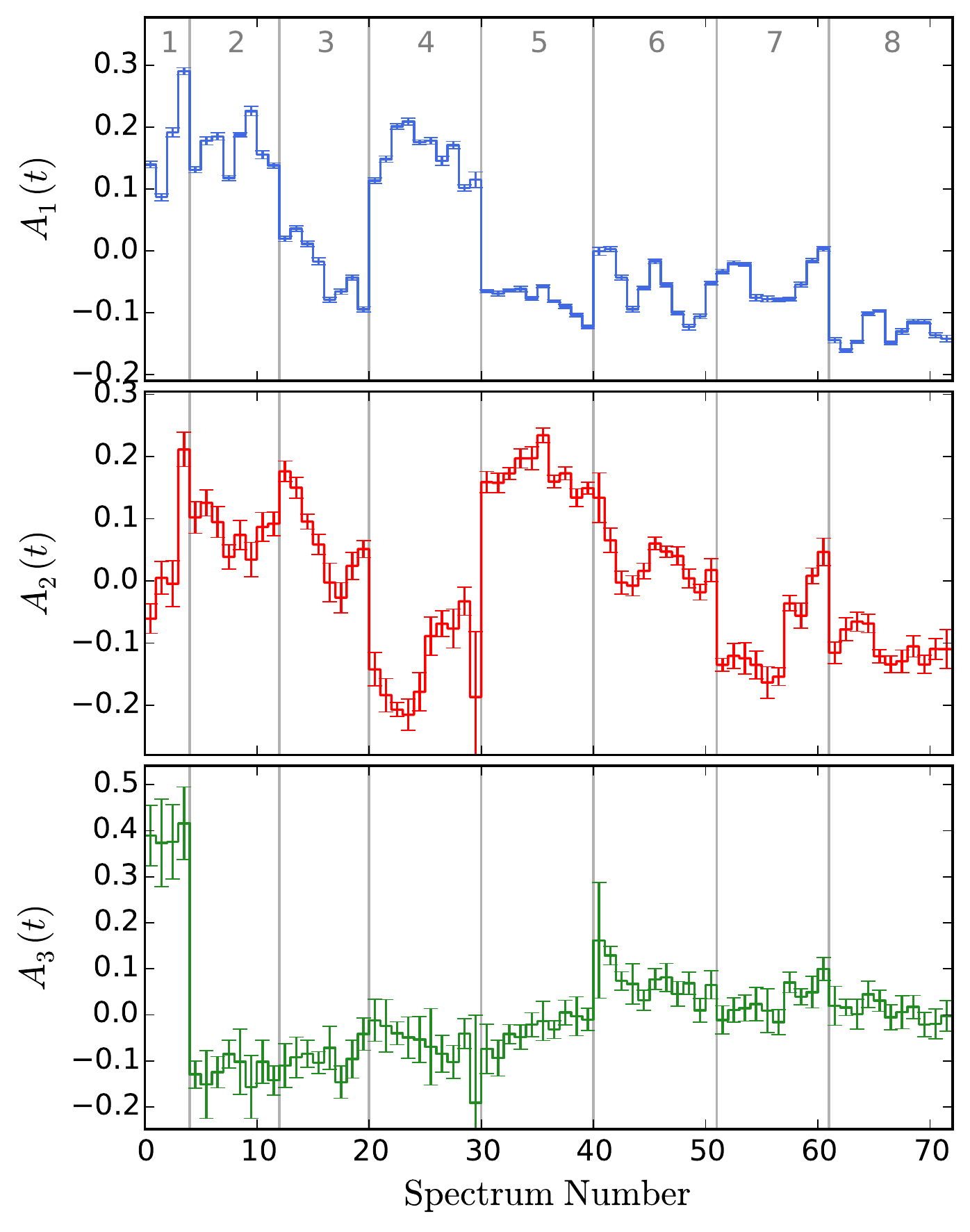}
\caption{Left: Component spectra for the first three PCs from the analysis of all \xmm\ EPIC-pn data of PDS 456. Right: Corresponding lightcurves. Vertical lines correspond to the different observations. Of particular interest are interval 1, where PC3 is extremely strong, and interval 4, where PC1 is high and PC2 is very low.}
\label{fig_epicpn_pcs}
\end{figure*}

These components are shown in Fig.~\ref{fig_epicpn_pcs}. The first component, PC1, is relatively flat, with a strong peak at $\sim9$~keV, corresponding to the energy of the UFO line, and a suppression of the variability at low energies, corresponding to the soft excess. The low energy suppression is frequently found in AGN \citep{Parker15_pcasample}, and indicates that the soft excess is less variable than the primary powerlaw continuum. This also means that the soft excess must be a real emission component, separate from the main continuum and less variable. While it is possible that the lack of variability could be due to a less variable reflection component \citep[where the variability is due to changes in the coronal size or geometry, see e.g.][]{Miniutti03, Miniutti04}, the absence of a corresponding iron line feature argues against this interpretation and suggests that the soft excess is due to a separate source of X-ray emission. The peak in variability at the UFO line indicates that the flux in this band is significantly more variable than the continuum, and that it is correlated with the continuum variability. As in the case of IRAS~13224-3809 \citep{Parker17_irasvariability}, this is most likely due to ionization of the outflowing gas by the increased source flux. 

PC2 is also fairly flat, with a strong anticorrelation between low and high energies. This is due to spectral pivoting - as the low energy flux rises, the high energy flux drops, and vice versa. Physically, this behaviour can be produced by changes in the photon index, or by small changes in the absorbing column of neutral gas along the line of sight to the source. We note that \citet{Nardini15} found evidence that the intrinsic photon index was variable during the 2013/14 \xmm /\nustar\ observing campaign ($\Gamma=2.1$--$2.6$).

The third component is more complex. There is a broad peak around 6~keV, and a soft excess feature, which are both anti-correlated with intermediate energies. This sort of qualitative variability has been seen in sources with strong reflection components in the past, but there are several factors that mean we cannot robustly attribute this variability to a specific physical mechanism (see \S~\ref{section_discussion}). Interestingly, this component is only strongly present in the first observation, dropping drastically before the second observation and never recovering to its original strength.

\begin{figure}
\includegraphics[width=\linewidth]{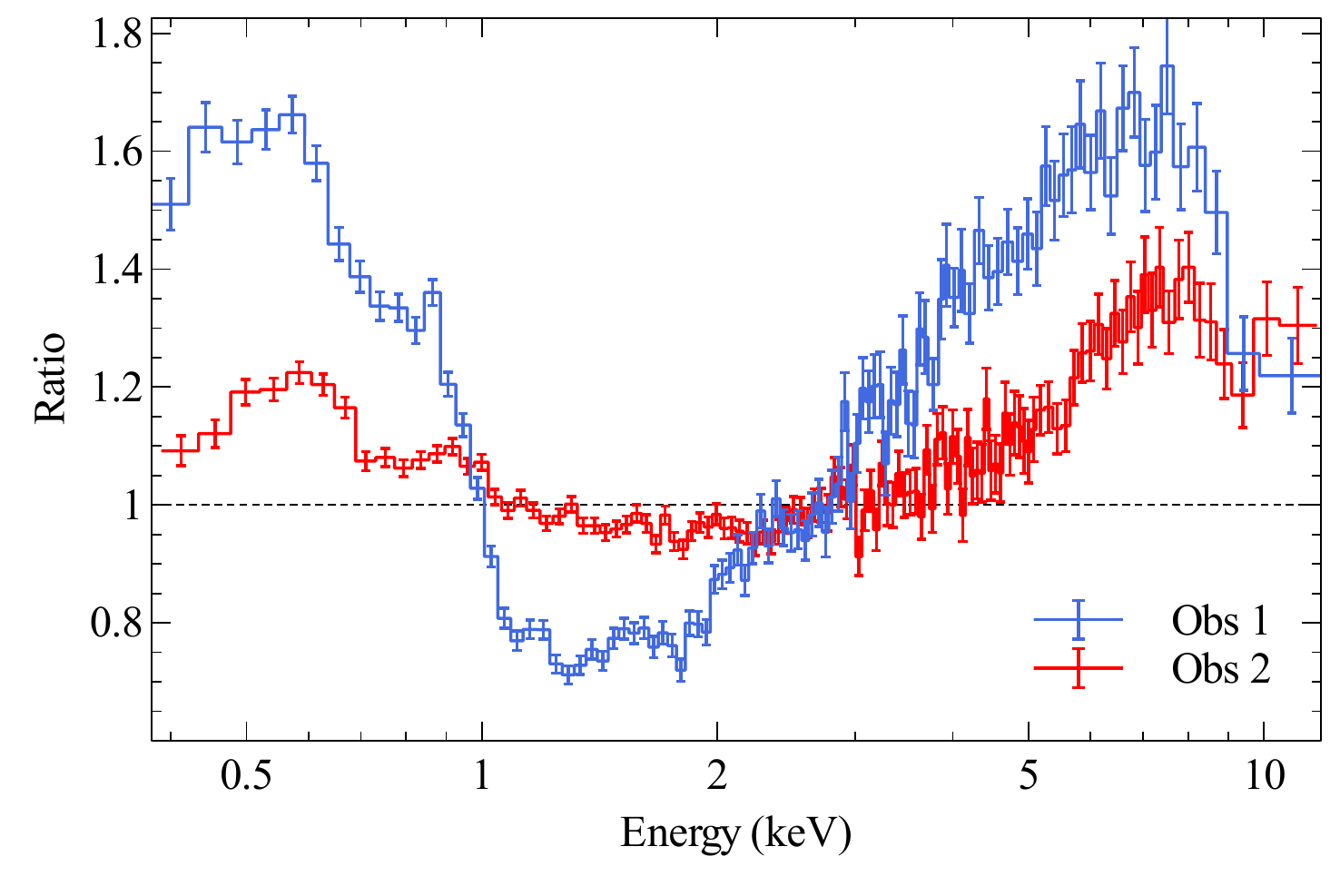}
\caption{Ratios of the EPIC-pn spectra from the first two \xmm\ observations of PDS~456 to a powerlaw, modified by Galactic absorption.}
\label{fig_ratio_comparison}
\end{figure}

To examine the changes between observations 1 and 2 in more detail, we examine the full spectrum of each observation. We fit the two spectra with a power-law and Galactic absorption from 0.3--10~keV (in the observer's frame) and plot the data/model ratio in Fig.~\ref{fig_ratio_comparison}. As expected from the spectral shape of PC3, in the first observation there is significant additional emission below 1~keV and from 3--9~keV. We discuss the nature of this variability more in \S~\ref{section_discussion}.

\subsection{Suzaku}

In contrast to the \xmm\ data, the spectral variability observed by \suzaku\ is dominated by variable cold absorption \citep{Matzeu16}. In particular, the three observations from a 2013 observing campaign show greatly increased absorption, with the flux below $\sim5$~keV dropping dramatically. 

\begin{figure*}
\centering
\includegraphics[width=0.4\linewidth]{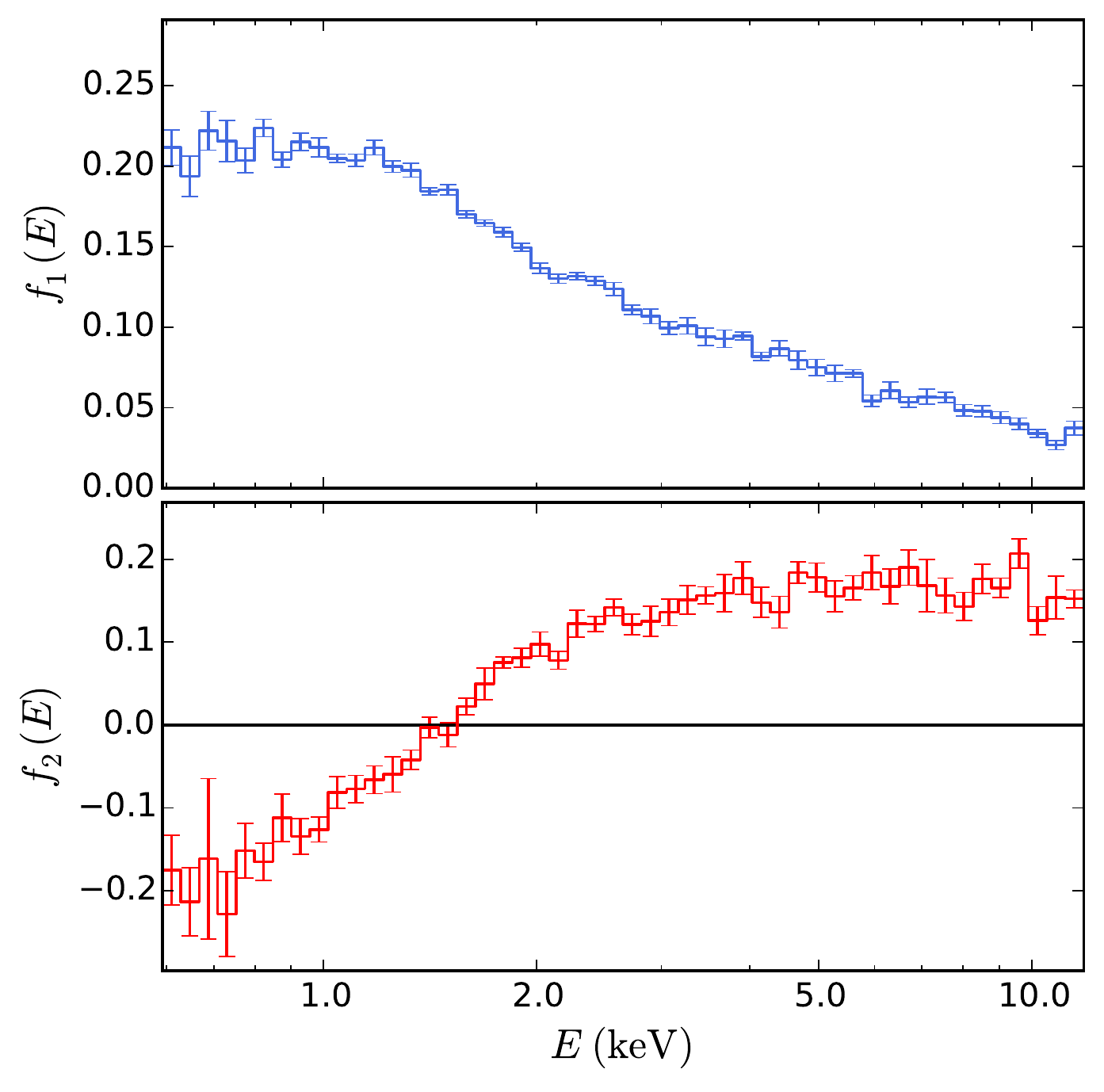}
\includegraphics[width=0.4\linewidth]{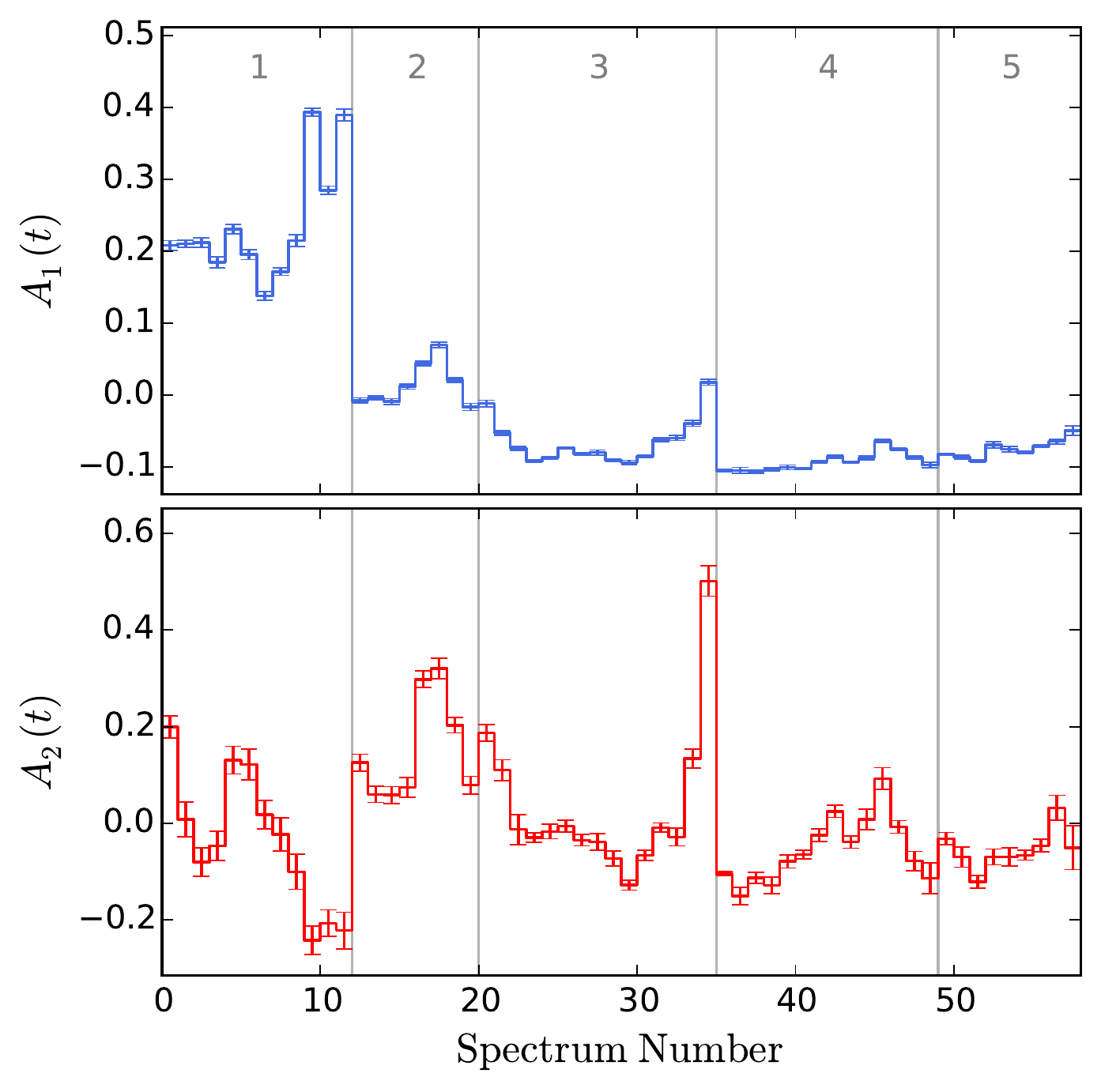}
\caption{First and second PCs from our analysis of the \suzaku\ data, and their corresponding lightcurves. The variability of the 1st PC is much softer than that found in the \xmm\ data, consistent with increased absorption variability, and neither component shows a significant UFO feature.}
\label{fig_suzakupcs}
\end{figure*}

The two significant PCs returned from this are shown in Fig.~\ref{fig_suzakupcs}, and are clearly showing a different variability pattern from those seen with \xmm , with a dominant soft component and a less variable hard component, which could potentially be closely related to the first two components from \xmm , but in reverse order (we discuss this further in \S~\ref{section_discussion}). PC1 is very soft, consistent with the components observed in sources dominated by absorption variability \citep[see e.g. NGC~4395 in][]{Parker15_pcasample}. This is mostly likely caused by a change in covering fraction or column density of the cold gas causing the absorption observed in the spectrum.

The second term is spectrally hard, and could potentially be caused by either intrinsic variability of the source or a second, higher column density, absorption component. Neither of these components have a strong UFO feature, despite the strong variability of the UFO feature seen in the 2013 \suzaku\ observations discussed by \citet{Matzeu16}. The combination of these two soft and hard components also reproduces the soft and hard flares seen in \citet{Matzeu17_flares}, at the end of the first and third observations.

\subsection{Simulations}
\label{sec_simulations}

To further investigate the UFO variability seen with \xmm , we perform simulations as in \citet{Parker15_pcasample}: by simulating a set of fake spectra based on simplified models and analysing them with PCA, we can compare to the results with real data and identify the likely causes of the observed variability.

We construct a simple model for the variable spectrum of PDS~456: a constant black body at low energies to mimic the less variable soft excess, a power-law varying in normalization for the continuum, and a Gaussian absorption line (\textsc{bbody+gabs*powerlaw} in \textsc{Xspec}). We consider three cases, where the absorption line is correlated, anti-correlated, and uncorrelated with the continuum flux. We vary the power-law normalization randomly between 0.5 and 2, and then fix the Gaussian normalization to be proportional, inversely proportional, or vary independently over the same range, respectively (for the power-law parameters used, this gives a comparable equivalent width of the Gaussian line to that observed).

\begin{figure}
\centering
\includegraphics[width=\linewidth]{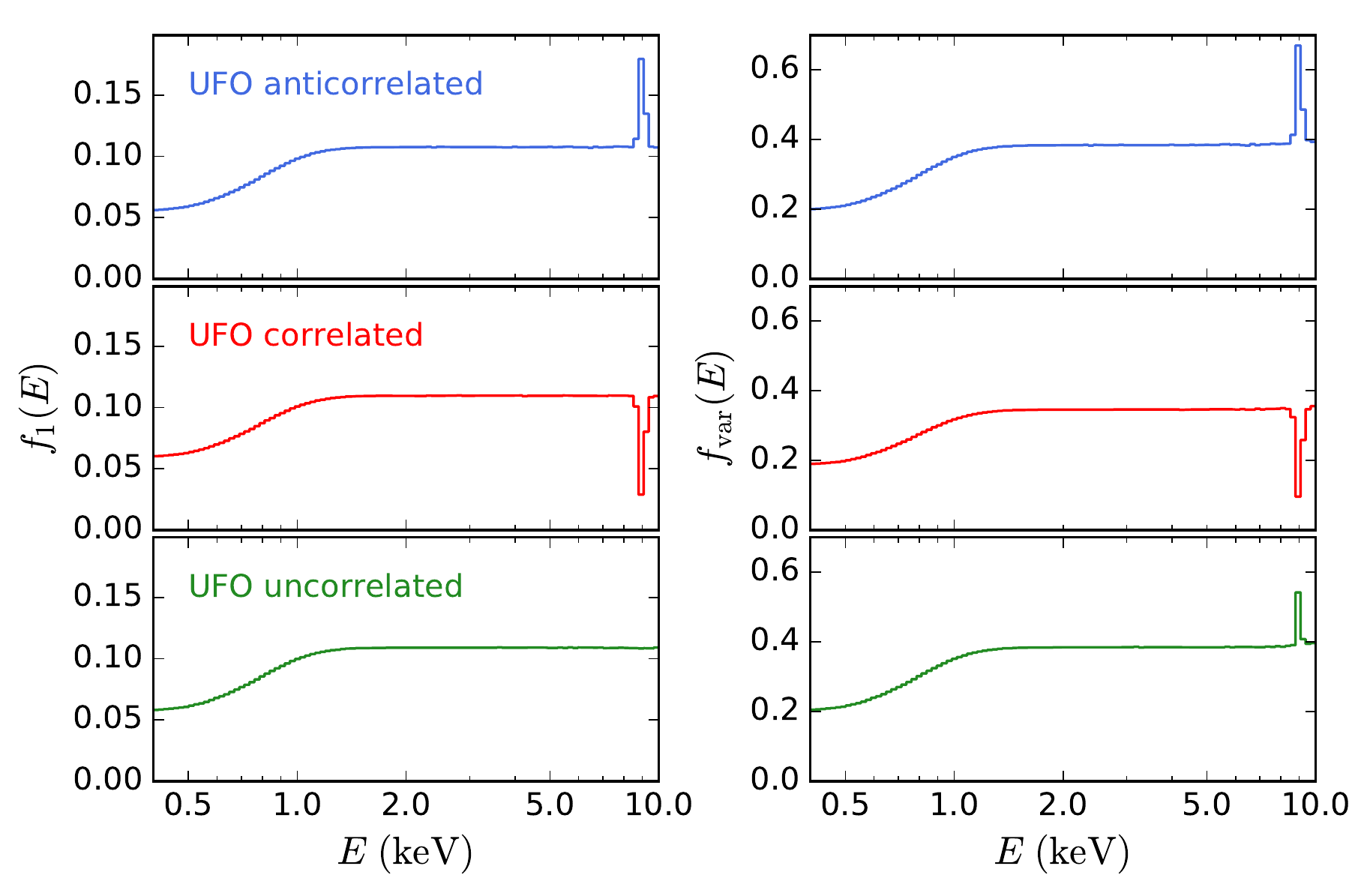}
\caption{A comparison of simulated PC (left) and RMS (right) spectra. We test three cases, where the equivalent width of the UFO line is anticorrelated, correlated, and uncorrelated with the continuum flux. PCA returns three qualitatively different spectra for the three simulations, whereas the RMS spectra for the uncorrelated and anticorrelated cases both show a spike in variability at the UFO energy.}
\label{fig_sims}
\end{figure}

The results of these simple simulations are shown in Fig.~\ref{fig_sims}. The only difference is at high energies, where the absorption line feature strongly depends on the source behaviour. If the equivalent width of the line is inversely proportional to the continuum flux, a strong positive feature is visible at 9~keV, as observed in the data. Alternatively, if the equivalent width is correlated with or independent of the continuum, the line feature is negative or absent, respectively. We also show in the right panels of this figure the corresponding $F_\mathrm{var}$ spectra, which are generally similar but do not qualitatively distinguish between anti-correlated and uncorrelated variability.

We also test the effect of allowing the velocity of the absorption line to respond to the source flux, instead of the equivalent width of the line responding. This could potentially mimic the effect of the equivalent width responding, by blue- or red-shifting the line into a different band as the flux changes. As this is not a type of variability that can be described by additive spectral components, PCA is not very sensitive to it.
For this, we vary the energy between 8 and 10~keV, proportionally to the log of the flux. This does produce a peak at the mean energy of the feature in PC1, but also gives a sharp drop above this feature. Additionally, it produces a series of components with peaks at different energies within the defined range, so that the feature can be reproduced. For the more moderate variability and smaller velocity shifts observed in the real data, the main effect of this will be to broaden the observed line width.
The typical energy shift ($\Delta E\sim\pm0.5$~keV) is well within the width of the line feature in the PCA spectrum, which stretches from 8--10~keV. It is likely that some correction components are produced to correct for energy shifts, but they presumably contribute little to the total variance and are lost in the noise components.

\section{Discussion}
\label{section_discussion}

\begin{figure}
\centering
\includegraphics[width=\linewidth]{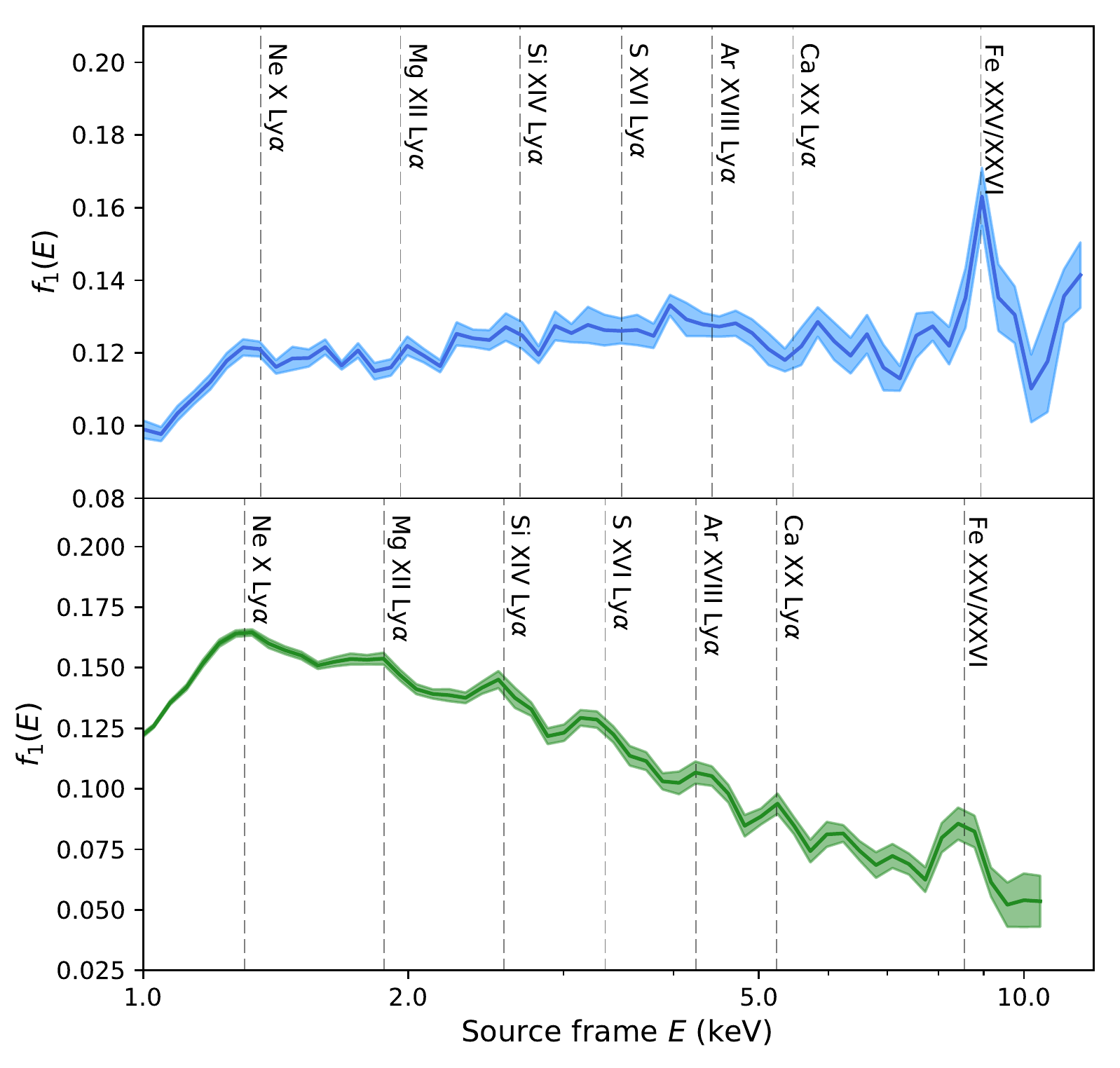}
\caption{A comparison of the \xmm\ PC1 spectrum of PDS~456 (top) and IRAS~13224-3809 (bottom). We label the Fe\textsc{xxv/xxvi} feature, and mark the energies corresponding to the strongest high ionization absorption features at lower energies. PDS~456 has a stronger, sharper Fe line, but there are no obvious corresponding features at lower energies. The rest-frame centroid energies are 8.9 and 8.6~keV, respectively. }
\label{fig_iras_comparison}
\end{figure}

The most striking aspect of our results is the detection of the 9~keV absorption feature in the \xmm\ data with PCA. The UFO line is significantly more variable than the continuum, and also inversely correlated with the continuum flux. This is very similar to the strong anticorrelation between the equivalent width of the UFO line and the continuum flux observed in IRAS~13224-3809 \citep{Parker17_nature}, and is most likely caused by the increased X-ray emission ionizing the wind, as discussed in \citet{Nardini15}.
In Fig.~\ref{fig_iras_comparison}, we show a comparison of PC1 from PDS~456 and IRAS~13224-3809. The Fe peak is sharper and stronger in PDS~456, but there are no significant lower energy features. The most obvious reason for this is that the ionization of the outflow in PDS~456 is higher, and the range of fluxes is much smaller, so the ionization does not drop low enough to produce the lower energy lines. Similarly, at lower fluxes in IRAS~13224-3809 the ionization is low enough that the Fe~\textsc{xxv} line starts to weaken again, which weakens the correlation between its equivalent width and the source flux, making the variability peak weaker in comparison to PDS~456. There are lower energy absorption features from relativistic outflowing gas at a lower ionization detected in the \xmm\ RGS data \citep{Reeves16}, but these are presumably due to different gas from that detected here, and are difficult to detect in the EPIC-pn data. A note of caution is that PDS~456 and IRAS~13224-3809 are sampled on very different effective timescales, due to the 2--3 orders of magnitude higher mass of PDS~456. This should not have a major impact on the UFO features, as the recombination time of the gas is very short in the case of IRAS~13224-3809, due to the high densities involved, and still much shorter than the dominant inter-observation variability in PDS~456.

The spectral shape of the \xmm\ PC3 is  very similar to that found in MCG--6-30-15 \citep{Parker14_mcg6}, 1H~0707-495 and others \citep{Parker15_pcasample}, which we identified as being due to relativistic reflection, with the enhanced soft excess and iron line emission caused by an increase in the reflection fraction. However, in this case there are some key differences which prevent us from confidently reaching the same conclusion. Firstly, and most importantly, this component is only seen in one observation, and is not strongly variable within that observation. In the other sources where this is observed, PC3 is present, and variable, in multiple observations, so we can be confident that it is not due to inter-observation variability, and that the soft excess and iron line features are correlated. Because we only see this component in one observation in PDS~456, we cannot be confident that the increase in soft flux and iron band flux are not coincidental. Secondly, the PC1 does not show the suppression around the iron line associated with relativistic reflection seen in other sources. This deficit is usually seen at the peak of the iron line, where the reflection component is less variable than the primary continuum (presumably due to light-bending effects), damping out the continuum variability. No such feature is visible in PC1 of PDS~456, although the soft excess damping is clearly evident.
We note that the 7~keV broad emission present in the spectrum has previously been interpreted as both relativistic reflection \citep[e.g.][]{Walton10} and scattered emission from the wind \citep[e.g.][]{Nardini15}. We note the recent detection of a reverberation lag in PDS~456 by \citet{Chiang17}, which is strong evidence for at least some relativistic reflection in the spectrum. We do not find clear evidence in favour of either interpretation using PCA, and further work is needed to understand how and if the two can be distinguished with this method in PDS~456.

\begin{figure}
\centering
\includegraphics[width=\linewidth]{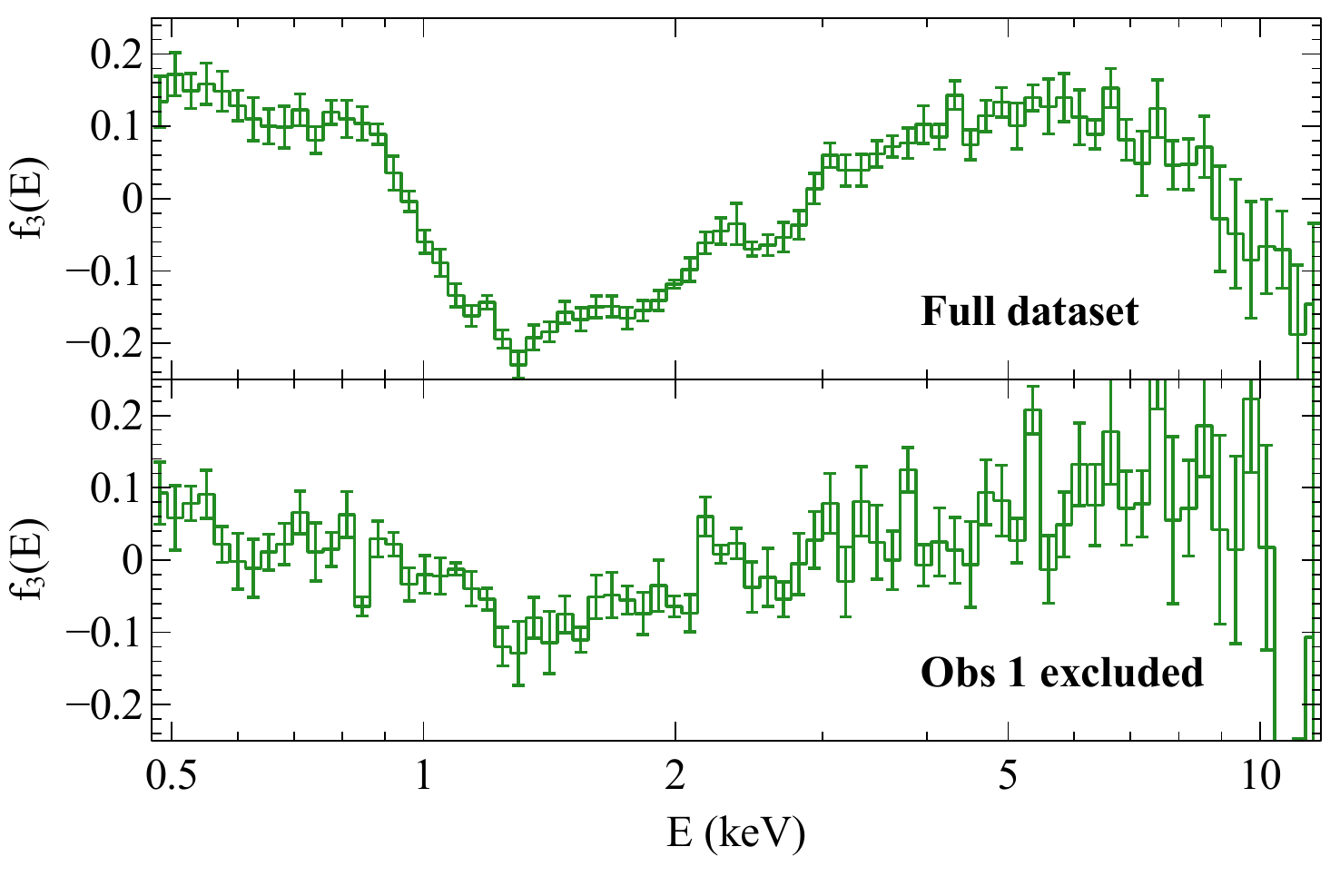}
\caption{Comparison of PC3 in the full dataset (top) and excluding the first observation, which dominates this component (bottom).}
\label{fig_PC3_comparison}
\end{figure}

There is some variability in PC3 after the first interval, but it is much weaker and we cannot rule out it being due to an entirely different mechanism. We show in Fig.~\ref{fig_PC3_comparison} a comparison of PC3 from the full dataset and with observation 1 removed, from which it is obvious how much observation 1 dominates the signal in this component. The noise increases dramatically when it is removed, and much of the structure is lost. This shape could potentially be due to relatively small changes in ionized absorption.

We previously used PCA to investigate the variability of PDS~456 as part of a sample of sources in \citet{Parker15_pcasample}. When that work was written, only the first three \xmm\ observations (from 2001 and 2007) were publicly available, and the results obtained differed significantly form those we find here. In particular, the variability of the first three observations is dominated by the large spectral changes between observations 1 and 2. This is largely associated with PC3 in this work, which is almost identical to PC2 from \citet{Parker15_pcasample}. However, the fractional variability is much higher for this component in that work: 9\%, compared to 3\% here. PC1 from \citet{Parker15_pcasample} is significantly different from that found in this work - instead of a flat power-law, suppressed at the soft excess and enhanced at the UFO line, the variability was suppressed at intermediate energies, and stronger around the energies of the iron band and soft excess, with no clear signal at the UFO energy. This is also due to the effect of the huge change between the first two observations.

\begin{figure}
\includegraphics[width=\linewidth]{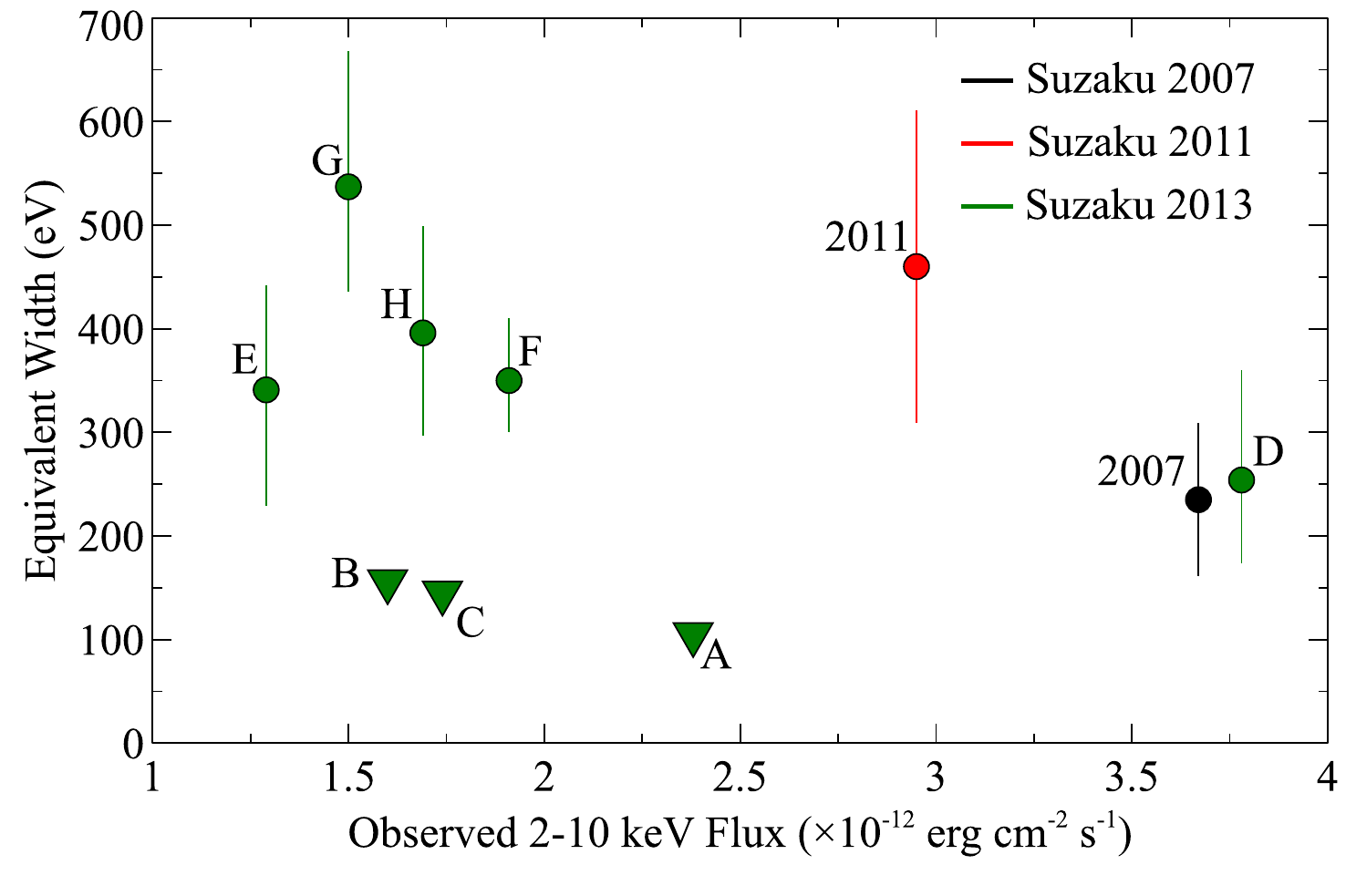}
\includegraphics[width=\linewidth]{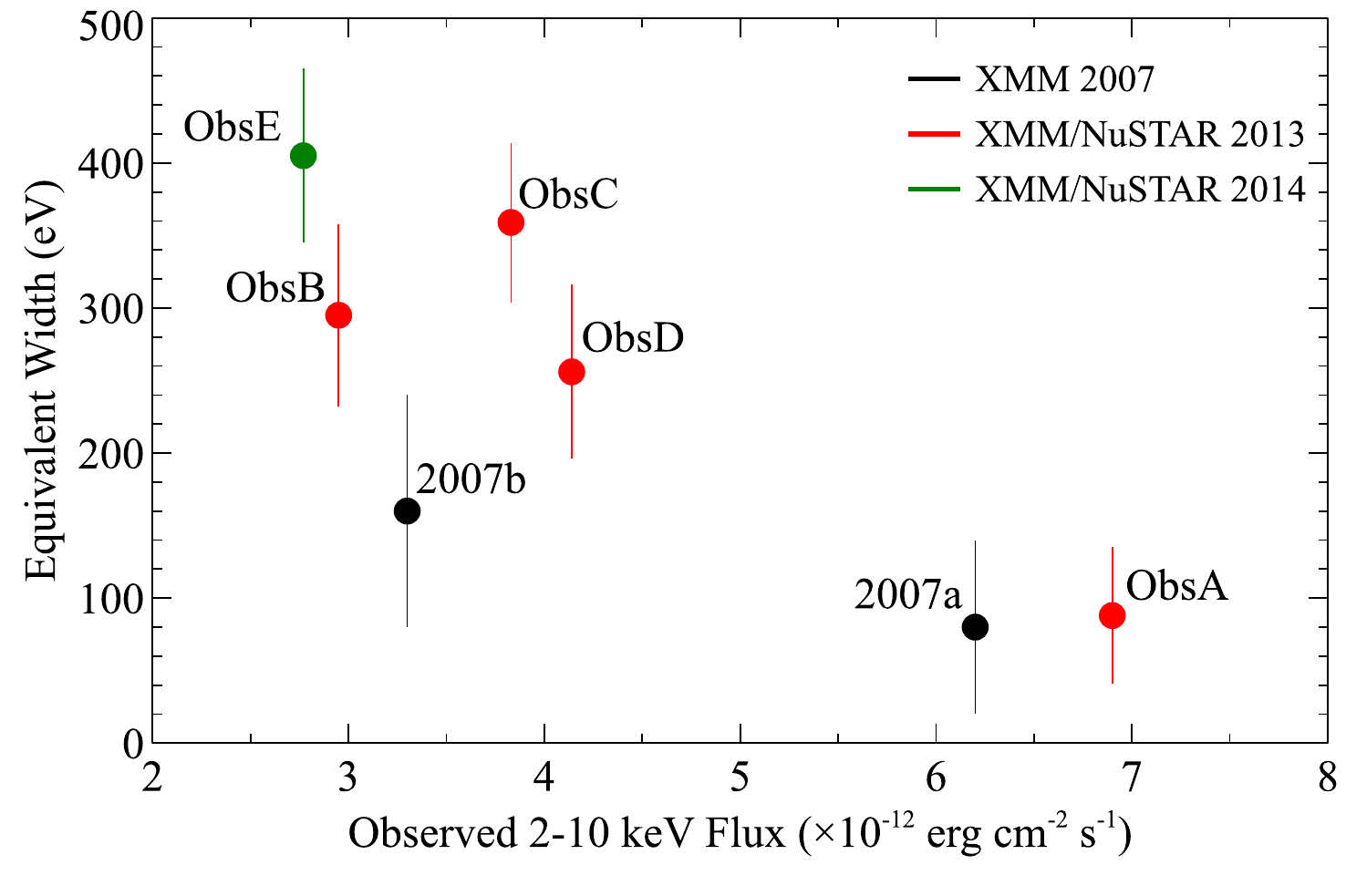}
\caption{Equivalent width of the Fe absorption line against 2--10~keV flux, corrected for Galactic absorption, for \suzaku\ \citep[top; A--H correspond to the spectra from][]{Matzeu16} and \xmm\ \citep[bottom; A--E correspond to the spectra from][]{Nardini15}. Triangles in the \suzaku\ plot are upper limits. The equivalent width appears to be anti-correlated with flux in the \xmm\ data, while no trend is evident in the \suzaku\ points.}
\label{fig_fluxplots}
\end{figure}

It is interesting that we do not find a strong UFO signature in either of the components from the Suzaku decomposition. \citet{Matzeu16} found extensive changes in the equivalent width of the 9~keV line between these observations, so we would expect to see some evidence of it in the variability. One possibility is that the line strength is correlated with the source flux (as in the \xmm\ data), but we cannot cleanly isolate that as a component due to the dominant low ionization absorption variability. Alternatively, the UFO strength could be correlated with \emph{both} the low ionization absorption column/covering fraction and the intrinsic flux, in such a way that our analysis does not find it to be correlated with either individually. These two components are (at least superficially) very similar to those found from simulations of partial covering absorption where the covering fraction and intrinsic flux both change \citep[][Fig.~12]{Parker15_pcasample}, which would argue for the intrinsic interpretation, but in this case we would expect to see the UFO feature, as in the \xmm\ data. Alternatively, the high energy variability could be caused by a higher column density absorption component, but in this case we would expect to see the 7~keV iron edge in PC2, as in the case of NGC~1365 \citep{Parker14_ngc1365}. One possibility is that both processes are driving the 2--10~keV variability, so PC2 describes the broad band variability and the narrow features (UFO line/Fe edge) are described by higher order components, which are lost in the noise.
In Fig.~\ref{fig_fluxplots} we show the equivalent width of the line against the source flux for the \suzaku\ and \xmm\ observations. It is clear that there is trend to lower equivalent widths at higher fluxes in the \xmm\ data, but there is no obvious pattern to the distribution of points taken from the \suzaku\ data. This is consistent with the scenario that this observing campaign is dominated by an absorption event, which obscures or scrambles the simple correlation with flux.

A weakness of this kind of analysis is that it is not sensitive to shifts in energy, as it assumes simple linear additive components. The velocity of the outflow in PDS~456 has been shown to be strongly correlated with the X-ray luminosity \citep{Matzeu17}, which could potentially impact our results. It could, for example, introduce a spurious correlation between the equivalent width and the flux if the line is blueshifted to a different energy band at higher fluxes. We explored this possibility in \S~\ref{sec_simulations}, and found that if this is the dominant form of variability on the timescale probed then it produces a qualitatively different series of PCs than observed. Additionally, the equivalent widths in Fig.~\ref{fig_fluxplots} take into account any changes in centroid energy, giving an independent check that the equivalent width is anticorrelated with flux. Energy shifts are clearly present in the data presented by \citeauthor{Matzeu17}, but they do not appear to be strongly affecting our results, or even being picked up by our analysis (except possibly as a broadening of the UFO feature). It is likely that there are higher order PCs which describe this variability, but they are lost in the noise.

\section{Conclusions}
\label{section_conclusions}
In this paper, we present the detection of the PDS~456 UFO using principal component analysis. This is the second such outflow to be picked up with this technique, which represents a promising method for detecting X-ray outflows in large datasets or rapidly variable, high signal data. Our main conclusions are summarised here:
\begin{itemize}
\item We find a strong peak in the variability spectrum of PDS~456 at the energy of the Fe absorption feature produced by the powerful outflow in this source.  This indicates that, as in the case of IRAS~13224-3809 \citep{Parker17_nature}, the ionization of the relativistic gas is responding to the source flux.
\item Unlike IRAS~13224-3809, we do not find any signatures of lower-energy (1--6~keV) absorption lines. This is presumably due to the higher ionization of the gas in PDS~456, which is a quasar rather than a NLS1, and the absence of a sufficiently large drop in flux in PDS~456, which means these ions cannot recombine.
\item There is a clear qualitative difference between the \xmm\ and \suzaku\ data. The variability observed with \xmm\ is dominated by the continuum, which the UFO line responds to. The \suzaku\ variability, on the other hand, is produced by neutral absorption, which does not show a clear relationship to the UFO feature.
\item We demonstrate using simulations that PCA can distinguish trivially between UFO features that are correlated, anticorrelated, and uncorrelated with the continuum. However, it is much harder to detect UFO features with PCA when they are uncorrelated with the continuum.
\end{itemize}

\section*{Acknowledgements}
We thank the referee, Chris Done, for detailed and constructive feedback.
MLP is supported by a European Space Agency (ESA) Research Fellowship.
MLP and ACF acknowledge support from the European Research Council through Advanced Grant on Feedback 340492. 
GAM acknowledge support from the Italian Space Agency (ASI INAF NuSTAR I/037/12/0) 
DJKB acknowledges financial support from the Science and Technology Facilities Council (STFC).

\bibliographystyle{mn2e}

\end{document}

%% file: journal_defs.tex
\long\def\symbolfootnote[#1]#2{\begingroup%
\def\thefootnote{\fnsymbol{footnote}}\footnote[#1]{#2}\endgroup} 
\def\araa{ARA\&A}
\def\apj{ApJ}
\def\aap{A\&A}
\def\mnras{MNRAS}
\def\nat{Nature}